\def\year{2022}\relax
\documentclass[letterpaper]{article}
\usepackage{aaai22}
\usepackage{times}
\usepackage{helvet}
\usepackage{courier}
\usepackage[hyphens]{url}
\usepackage{graphicx}
\usepackage{natbib}
\usepackage{caption}
\DeclareCaptionStyle{ruled}{labelfont=normalfont,labelsep=colon,strut=off} 
\usepackage{algorithm}
\usepackage{algorithmic}

\usepackage{newfloat}
\usepackage{listings}
\floatstyle{ruled}
\newfloat{listing}{tb}{lst}{}
\floatname{listing}{Listing}

\usepackage{amsmath,amssymb,amsfonts}
\usepackage{algorithmic}
\usepackage{textcomp}
\usepackage{parskip}
\usepackage{amsmath}
\usepackage{tabularx}
\usepackage[title]{appendix}

\begin{document}
%
\title{Transformer-based Value Function Decomposition for Cooperative Multi-agent Reinforcement Learning in StarCraft}

 \author{Muhammad Junaid Khan, Syed Hammad Ahmed, Gita Sukthankar}
 \affiliations{
        Department of Computer Science\\
        University of Central Florida\\
        Orlando, FL US \\
	\{junaid\_k, hammad.ahmed\}@knights.ucf.edu, gitars@eecs.ucf.edu
}
\maketitle
\begin{abstract}
The StarCraft II Multi-Agent Challenge (SMAC) was created to be a challenging benchmark problem for cooperative multi-agent reinforcement learning (MARL).  SMAC focuses exclusively on the problem of StarCraft micromanagement and assumes that each unit is controlled individually by a learning agent that acts independently and only possesses local information; centralized training is assumed to occur with decentralized execution (CTDE).  To perform well in SMAC, MARL algorithms must handle the dual problems of multi-agent credit assignment and joint action evaluation.

This paper introduces a new architecture TransMix, a transformer-based joint action-value mixing network which we show to be efficient and scalable as compared to the other state-of-the-art cooperative MARL solutions. TransMix leverages the ability of transformers to learn a richer mixing function for combining the agents' individual value functions.  It achieves comparable performance to previous work on easy SMAC scenarios and outperforms other techniques on hard scenarios, as well as scenarios that are corrupted with Gaussian noise to simulate fog of war. 
\end{abstract}
\setlength{\parindent}{12pt}
\section{Introduction}


StarCraft poses many exciting research challenges for AI agents~\cite{hc_feat1}, stemming from both the myriad \textit{macromanagement} (strategic gameplay choices related to production and expansion) and \textit{micromanagement} (tactics related to movement and targeting) tasks that occur during full gameplay.  Even though humans typically play StarCraft in a centralized way in which the single human player controls all the units, the StarCraft II Multi-agent Challenge testbed~\cite{smac} treats engagements as a cooperative multi-agent problem in which each AI agent controls a single unit and has limited visibility.  SMAC provides a set of battle scenarios as benchmarks for cooperative multi-agent reinforcement learning (MARL), unlike PySC2 \cite{sc2le} which is designed for a single learning agent controlling all the units.  During execution, each SMAC agent conditions its decisions on the partially observable limited area of view; enemy units do not have the opportunity to adapt and are controlled by the built-in heuristic controller. Scenarios include the initial positions, count, and types of units, as well as terrain feature information.  Some of the SMAC scenarios are designed to encourage learning agents to acquire a \textit{micro-trick} such as kiting or focusing fire.  A scenario can be symmetric or asymmetric, based on the unit counts of allies and enemies, and homogeneous or heterogeneous according to the unit types of each side.\par 


Multi-agent reinforcement learning is significantly more challenging than single agent tasks due to the problem of \textit{credit assignment}: it's difficult for agents to determine whether it was its own action selection that yielded rewards or another agent's choices. A na{\"\i}ve solution is to concatenate each agent's individual state-action spaces and treat it as a single-agent problem \cite{centralized2}.  Both learning and execution are centralized and a single joint reward is learnt which cannot be decomposed into independent agents' contributions. The primary concern with this approach is the curse of dimensionality, as it creates a drastic growth in the total number of unique global states resulting in exponential space and time complexity. 

With a decentralized paradigm, agents independently condition only on their local observations without inter-agent communication and without a joint action-value function \cite{independent2}. Due to its decentralized nature, parallelism can be exploited, leading to faster learning but with no convergence guarantees. These limitations can be addressed by centralizing the learning and decentralizing each agent's execution --- known as Centralized Training with Decentralized Execution (CTDE)~\cite{b1}. 

The release of the SMAC environment has spurred a burst of innovation in multi-agent value learning algorithms, such as QMIX \cite{qmix}, QPLEX \cite{qplex}, QTran \cite{qtran}, and Qatten \cite{qatten}.  A key distinction between these algorithms is the credit assignment process which factorizes joint action-values into individual action-values for every agent.  To do this many MARL techniques make limiting assumptions about the value function to facilitate the learning process.  Adhering to the IGM (Individual-Global-Max) principle~\cite{qtran} avoids incompatible agent policies: an action selected using the joint action-value function should be equivalent to the greedy action selections of individual agents.

This paper introduces a transformer-based mixing approach, TransMix\footnote{Code is available at: https://github.com/junaiddk/transmix}, for  joint action-value learning in cooperative MARL. TransMix is able to extract global as well as local contextual interaction amongst individual agent Q-values, agent histories, and global state information, and can deal with longer time horizons which helps it achieve better performance on the hard SMAC scenarios, as well as scenarios that are corrupted with Gaussian noise. The next section presents background on the mathematical notation used throughout the rest of the paper.

 


\section{Background}
\textbf{Decentralized Partially Observable MDP (Dec-POMDP)}: MARL tasks are typically modeled as a Dec-POMDP \cite{b1}, described by the tuple $\mathbf{ \mathcal{M} = \mathcal{<} N, S, \mathcal{A}, P, r,\Omega, O, n, \gamma \mathcal{>} }$ where $i \in N = \{1, 2, 3, ...\}$ is the set of agents while $s \in S$ represents the true or global state in the environment. For every time step, agent $i \in N$ selects an action $a_i \in \mathcal{A} \equiv \mathcal{A}^n $ based on state $s$. When the joint action $a$ is executed, a joint reward $r(s, a)$ is received, resulting in a transition to the next state $s'$ based on a transition probability function $P(s^{'} | s, a)$ with a discount factor $\gamma [0, 1)$. 

To make it a partially observable process, each agent $i$ receives an observation $o_i \in \Omega$ based on the observation probability function $O(o_i | s, a_i)$. In addition, not only does each agent maintain an action-observation history function $\tau_i \in \tau \equiv (\Omega \times \mathcal{A}^*)$, it conditions its stochastic policy $\pi_i(a_i | \tau_i) $ on this history as well. The agents seek to achieve a joint policy $\Pi$ that maximizes the joint value function $\mathcal{V}^\pi(s)$ as well as joint action-value function $\mathcal{Q}^\pi(s, a)$.

\textbf{Deep Q-Learning}:  Q-learning algorithms maximize the action value function $\mathcal{Q}^{*}(s, a) = r(s, a) + \gamma \mathbb{E}_s{'}[\max_{a}^{'}\mathcal{Q}^{*}(s^{'}, a^{'})]$.  In deep Q-learning, this function is learned using a deep neural network with parameters $\mathcal{\theta}$. This idea of deep Q-networks was first introduced by \citeauthor{dqn} (\citeyear{dqn}) who used replay memory to store transition tuples of the form $(s, a, r, s^{'})$, where $r$ is the reward of taking action $a$ resulting in a state transition from $s$ to $s^{'}$. The network parameters, $\theta$, are optimized by sampling batches from this replay memory in order to eliminate correlations in the tuple sequence.      

Standard DQN-based approaches utilize the temporal difference (TD) loss to optimize the network, given by the equation
\begin{equation}
    \mathcal{L}(\theta) = \sum_{i=1}^{b}[(y_{i}^{DQN} - Q(s, a; \theta)^{2}]
    \label{eq1}
\end{equation}
where $y_{i}^{DQN}$ represents the target network which is updated at regular intervals (rather than at each iteration) and $b$ is the batch size sampled from the replay buffer.

In order to deal with the problem of partial observability, a recurrent version of the Q-network (DQRN) \cite{dqrn}, has been used where a recurrent neural network (RNN) based agent learns temporal information after each state transition. In this case, the replay memory stores the joint action-observation history tuple $(\tau, a, r, \tau^{'})$ and the Q-value is calculated using $Q(\tau, a; \theta)$ instead of $Q(s, a; \theta)$.


\textbf{Training and Execution}: Recent MARL value factorization techniques benefit from the usage of a Centralized Training and Decentralized Execution (CTDE) paradigm \cite{b1}.  \citeauthor{independent2}  (\citeyear{independent2}) demonstrated that by sharing individual observations and policies, independent agents learn a cooperative task substantially quicker at the cost of communication and space overhead. Without cooperation, the independent Q-learners may not converge even if exhaustive exploration is assumed. The Centralized Training with Decentralized Execution paradigm addressed the issues with independent learning agents \cite{independent1, independent2} and fully-centralized learning approaches \cite{centralized1, centralized2} by avoiding misleading agent reward assignments, and eliminating the need for combined action and observation spaces.

In CTDE, agents are trained in a central setting where each agent has access to global state information. At training time, each agent aims to maximize its own action-value function which leads to the maximization of the team's joint action-value function. At execution time each agent selects its action based on its own learned action-value functions. The Individual-Global Max (IGM) concept was introduced by \citeauthor{qtran} (\citeyear{qtran}) who state that the optimal joint actions of agents are dependent upon optimal actions of individual agents i.e.,

\begin{equation}
    argmax_{\boldsymbol{a}} Q_{tot}(\boldsymbol{\tau, a}) = \begin{pmatrix}
            argmax_{a_{1}} Q_1 (\tau_1, a_1) \\
            argmax_{a_{2}} Q_2 (\tau_2, a_2) \\
            \vdots \\
             argmax_{a_{N}} Q_N (\tau_N, a_N) \\
        \end{pmatrix}
        \label{eq2}
\end{equation}

Value decomposition networks~\cite{vdn} calculate $Q_{tot}$ by summing up the individual $Q_{i}$ of each agents while QMIX~\cite{qmix} applies a monotonicity constraint.  These simplifications to the factorization process facilitate learning but are unable to represent some classes of joint action-value functions. Our research attempts to address this problem through the usage of a more complex transformer-based mixing function.


\section{Related Work}



Playing a full game of StarCraft requires creating an AI that can rapidly make both strategic (macromanagement) and tactical (micromanagement) decisions against opponents shrouded by fog of war on a large changing map (see \citeauthor{hc_feat1} (\citeyear{hc_feat1}) for a comprehensive overview of StarCraft research challenges).  Much of the recent research on reinforcement learning in StarCraft II has been conducted on micromanagement scenarios executed using either the StarCraft II Learning Environment (SC2LE)~\cite{sc2le} or the StarCraft II Multi-agent Challenge testbed~\cite{smac}, which is built on top of SC2LE. SMAC models battles as being executed by multiple cooperative agents.  Rather than tackling the problem of learning policies for an entire game which requires significant computational power (e.g., such as what was used for DeepMind's AlphaStar system~\cite{sc2deepmind}), SMAC evaluates the ability of cooperative agents to learn micromanagement policies for a single battle.  The PyMARL framework was also released as part of SMAC to facilitate the development and performance comparisons of newly proposed RL algorithms with other popular implementations \cite{vdn, qmix, wqmix, qtran, qplex}.  The next section provides an overview of recent work on RL for SMAC.
    \subsection{MARL in StarCraft} 
   
Value decomposition networks (VDN) \cite{vdn} introduced the seminal idea of learning a unified reward value and factorizing it into agent-wise reward values; the algorithm assumes that the joint value function can be additively decomposed into individual value functions for each agent.  Since VDN only considers linear value functions, it does not perform well for complex benchmarks like the SMAC scenarios that have many agent interdependencies \cite{smac}.  Also unlike later MARL algorithms, VDN does not leverage global state information.

In QMIX, \citeauthor{qmix} (\citeyear{qmix}), improved on VDN by incorporating global state information during training and supporting a broader range of non-linear value functions. The QMIX architecture includes a mixing network conditioned on the state information which improves the representational complexity of the value functions.  The monotonicity of the joint value function is ensured through the usage of positive weights and individual agent networks conditioned on local observations only.   However QMIX’s loss function tries to minimize loss across all joint actions for every state; this can result in incorrect argmax selections when decomposing non-monotonic functions. Weighted QMIX \cite{wqmix} fixes some issues with the original architecture by assigning weights to the projected joint action-values instead of using equal weights as was done in the original QMIX paper; these weight assignments can either be done centrally or optimistically.

\citeauthor{qtran} (\citeyear{qtran}) identified some limitations with the usage of additivity (VDN) and monotonicity (QMIX) constraints.  They emphasize that an effective factorization of optimal joint action-values must instead satisfy the IGM (Individual-Global-Max) property such that the best joint action-value is the aggregation of individual agents' optimal action-values. Their algorithm QTRAN uses an additional state-value network to compete a scalar value representing the state. This value helps minimize the difference in the total joint reward and the sum of each agent's individual rewards, and therefore supports a variety of non-monotonic joint action-value functions. QPLEX \cite{qplex} ensures compliance to the IGM principle by introducing an advantage-based function realized with a duplex dueling architecture \cite{duelingDQN}; it is one of the best performing SMAC algorithms.  In this paper, we benchmark our work against both QMIX and QPLEX.
\begin{figure*}[ht!]
    \centering
    \includegraphics[width=1\textwidth]{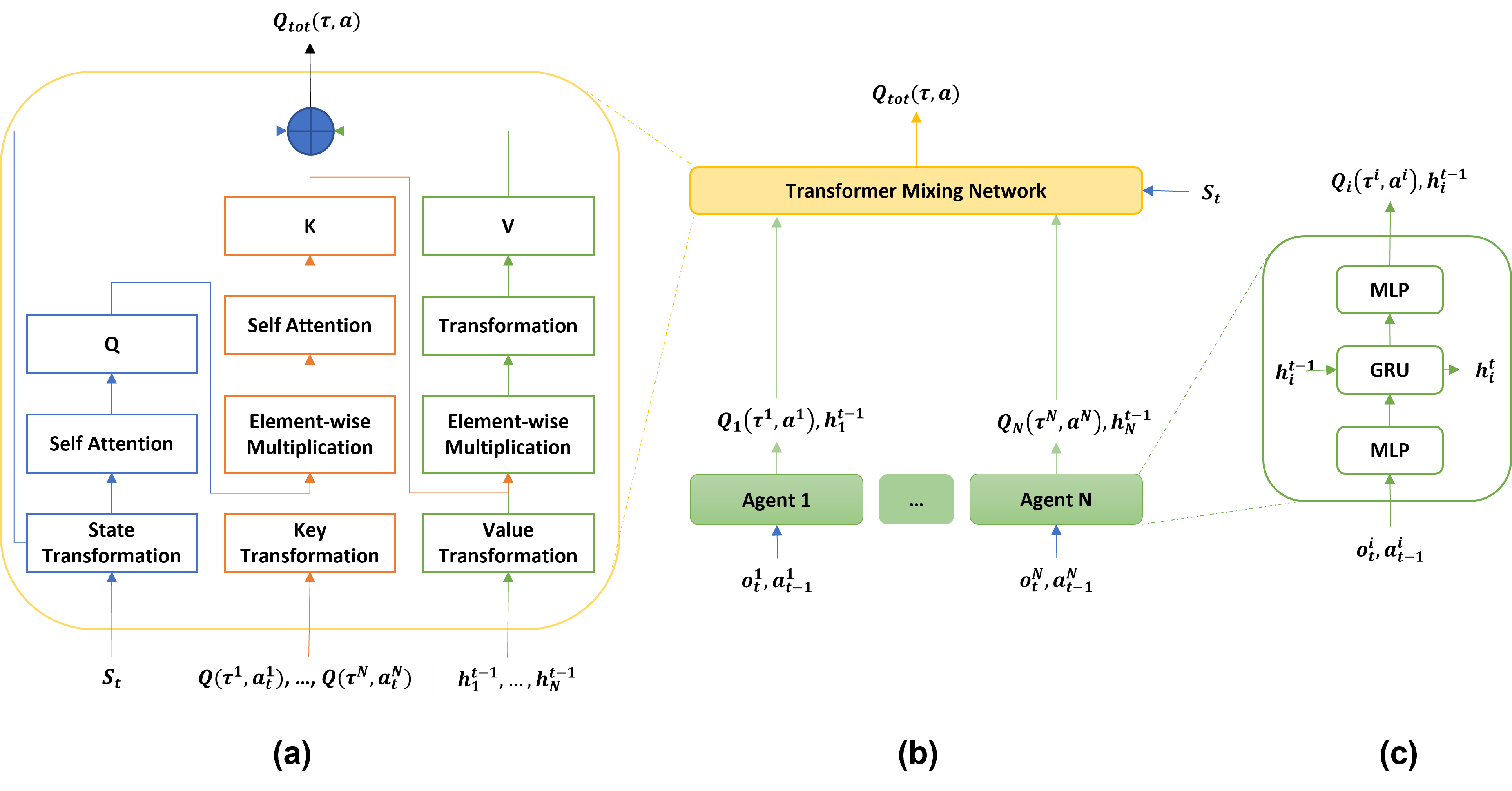}
    \caption{Complete architecture of TransMix. (a) is the transformer encoder. (b) is the overall architecture of the TransMix. (c) is the GRU-based agent network. A stack of 2 to 6 transformer encoder layers are used based on the complexity of the task.}
    \label{fig: TransMix Architecture}
\end{figure*}

 As the number of agents in a multi-agent problem increases, the role of context assumes greater importance since the agent must consider the states of other agents, but only a small subset of the other agents are relevant to the task on hand. To overcome this problem, \citeauthor{qatten} (\citeyear{qatten}) proposed incorporating a multi-headed attention module into the mixing network used for joint Q-value estimation. Unlike convolutional and recurrent architectures, attention mechanisms can model far-flung dependencies in input and output sequences~\cite{transformer}. Instead of proposing enhancements to the mixing of agent network structures, \citeauthor{apiboost} (\citeyear{apiboost}) sought to scale up to larger multi-agent problems by improving input sample efficiency.  They realized that in a system with a small set of cooperating agents, sampling all permutations of their ordered representations is redundant, hence the training sample number can be reduced by forcing agent-ordering to be permutation invariant.


Our work leverages the transformer model which relies exclusively on attention to learn data dependencies; also it can be designed to support permutation invariance.  \citeauthor{junaid0} (\citeyear{junaid0}) demonstrated promising results on the usage of transformer networks for macromanagement task prediction in StarCraft II; however, their work was done on a static dataset using supervised learning. Unfortunately the basic transformer architecture has been shown to be unstable with the shifting RL objective function~\cite{parisotto2020stabilizing} so embedding transformers into an online RL agent requires careful training. Fine-tuning an offline-learnt transformer with online RL training can improve both overall agent performance and sample efficiency~\cite{decisionTransformer}. 


\section{Method}
This paper introduces our transformer-based value decomposition and mixing network, TransMix. Our transformer design is inspired by the approach of Fastformer \cite{fastformerAA}.  Each agent is represented by a GRU-based DQRN network. At every time step, agents receive observations $o^i_t \in \Omega$ and previous action $a^i_{t-1} \in \mathcal{A}$. Based on $o^i_t$ and $a^i_{t-1}$ the agent network estimates the individual $Q_i (\tau^{i}, a^{i})$ where $\tau$ is the action-observation history and selects the next actions for each agent by following an $\epsilon$-greedy policy.  

The transformer encoder consists of 2 to 6 transformer layers. The number of transformer layers or depth is based on the complexity of the task at hand. Once a batch of data is ready, it is fed to the transformer which ingests individual $Q_i$, action-observation histories $h_i^{t}$ and global states $S_t$.

We calculate the query vector, \textbf{Q}, from the global states by applying linear transformation and self-attention. This operation ensures that the most important states get attention and contribute more towards the global context in learning the $Q_{tot}$. Similarly, we apply a separate linear transformation to individual $Q_i$ and the resultant is then multiplied element-wise with Q vector which gives a global context aware key vector between $S_t$ and $Q_i$. Next, we apply  self-attention to extract the most relevant information in the form of global \textbf{K} vector. This K vector is then element-wise multiplied with transformed $h_i^{t}$ that goes through further linear transformation to produce global context aware \textbf{V} vector. Lastly, the transformed states and V vectors are aggregated together to estimate the $Q_{tot}$. The complete TransMix architecture is presented in Fig. \ref{fig: TransMix Architecture}.

Following the design guidelines of Fastformer, we also make use of additive self-attention for calculating \textbf{K} and \textbf{Q} vectors rather than the standard self-attention proposed by the vanilla transformer design \cite{transformer}. The benefits of additive self-attention are many fold. Firstly this attention mechanism is rather light weight. Second, it reduces the total complexity of transformer encoder. Finally, it improves the inference time of the transformer as well.

The matrices \textbf{K}, \textbf{V}, and \textbf{Q} $\in \mathbb{R}^{\mathcal{N} \times d}$, where $\mathcal{N}$ $\in \mathbb{R}^{512}$ is the sequence length or the embedded dimension of the transformation layers and $d \in \mathbb{R}^{2048}$ is the hidden dimension. We also utilize 4 attention heads for our TransMix network. The linear transformations with sequence length or hidden dimension $\mathbb{R}^{512}$ ensure that we can provide a fixed length input to transformer layers which helps to keep the complexity and computation under control. 

We train TransMix in online data collection settings in an end-to-end fashion. The network is optimized with an Adam optimizer and standard TD loss (Equation \ref{eq1}). The learning rate is set to $0.001$ for most of the SMAC scenarios with $\beta_1 = 0.9$ and $\beta_2= 0.999$, and batch size is set to 96. We also utilize a skip connection that consists of concatenated $Q_i$ and $h_i^t$ which is passed through a bottleneck linear transformation. This bottleneck skip connection provides training stability and helps the network converge while reducing the dimensionality of the concatenated matrices.  Unlike many other methods, our approach does not use hypernetwork generated weights extracted from global states.  This makes our approach less dependent on global states compared to others. 
\begingroup
\renewcommand{\arraystretch}{1.25}
\begin{table*}[ht]
\centering

\begin{tabularx}{0.8\textwidth} { 
  >{\raggedright\arraybackslash}X 
  | >{\centering\arraybackslash}X 
  | >{\centering\arraybackslash}X 
  | >{\centering\arraybackslash}X
  | >{\centering\arraybackslash}X}
 
 \centering{\textbf{Maps}} & {\textbf{Difficulty}} & {\textbf{TransMix}} & {\textbf{QMIX}} & {\textbf{QPLEX}}\\  
 \hline
 \hline
 2m\_vs\_1z & Easy & 99. 7 & 95.3 & \textbf{100} \\
 3m & Easy & \textbf{100} & 96.9 & \textbf{100} \\
 8m & Easy & \textbf{100} & 97.7 & \textbf{100} \\
 2s3z & Easy & \textbf{100} & 88.25 & \textbf{100} \\
 1c3s5z & Easy & \textbf{97.62} & 93.37 & 97.2 \\
 MMM & Easy & \textbf{100} & 95.3 & 96.9 \\
 bane\_vs\_bane & Easy & \textbf{100} & 89.21 & 98.47 \\
 \hline
 3s\_vs\_5z & Hard & 95.91 & 88.7 & \textbf{99.1} \\
 3s5z & Hard & \textbf{96.68} & 88.3 & 95.9 \\
 5m\_vs\_6m & Hard & \textbf{77.41} & 69.2 & 73.4 \\
 8m\_vs\_9m & Hard & \textbf{96.88} & 92.2 & 87.5 \\
 2c\_vs\_64zg & Hard & \textbf{92.62} & 84.38 & 91.2 \\
 10m\_vs\_11m & Hard & \textbf{91.77} & 89.2 & 90.12 \\
 \hline
 
\end{tabularx}
\caption{Median win rate on SMAC maps. We train our network for 2M training timesteps on each map. TransMix outperforms QMIX and QPLEX on most hard SMAC scenarios and ties on easy scenarios.}
\label{table:1}
\end{table*}
\endgroup
\section{Results}
We evaluate TransMix on both easy and hard scenarios from the SMAC~\cite{smac} challenge benchmark.  In SMAC, the focus is on micromanagement tasks in SC2 such as unit battles. During the battles, we train our RL agents in an online fashion while the opponents are controlled by the built-in AI.
A battle is considered as ``win'' if the ally unit kills an opponent unit. On the other hand a ``loss" occurs if the opponent kills an ally unit or the maximum number of episodes are reached. When an RL agent damages the opponent, a reward is received proportional to damage done. Similarly, a reward of 10 is received for killing an opponent's agent; winning the battle accumulates a reward of 200. The details of SMAC scenarios are provided in Table \ref{table:3}. 

\subsection{Training and Evaluation}
We follow the same evaluation metric proposed by \citeauthor{qmix} (\citeyear{qmix}). 
For each SMAC map, the experiment is repeated 5 times. We train the model for 2M timesteps on all the scenarios with a replay buffer capacity of 5000 episodes; linear $\epsilon$ annealing from 1.0 to 0.05 is performed over 50k steps for easy maps while for the hard maps the range varies from 250k to 500k steps. After every 10k timesteps, we pause the training and test our method for 20 test episodes in a decentralized fashion. Moreover, after every 200 episodes, we update our target network parameters. 

The main results are summarized in Table \ref{table:1}. Since our approach is based on transformer, which is highly scalable, we leverage the built-in parallel episode runner environment for training and testing which substantially reduces the training time for TransMix. We benchmark our work against both QMIX~\cite{qmix} and QPLEX~\cite{qplex} which provide a good illustration of the performance of our method.  QMIX is one of the oldest methods to be successful at SMAC, whereas QPLEX is a recent top performer that incorporates all the value function decomposition improvements proposed by earlier authors.  Our method always outperforms QMIX and performs better than QPLEX in the majority of the hard scenarios.

\subsection{Effects of Noisy States}
\begin{figure*}[ht!]
    \centering
    \includegraphics[width=1\textwidth]{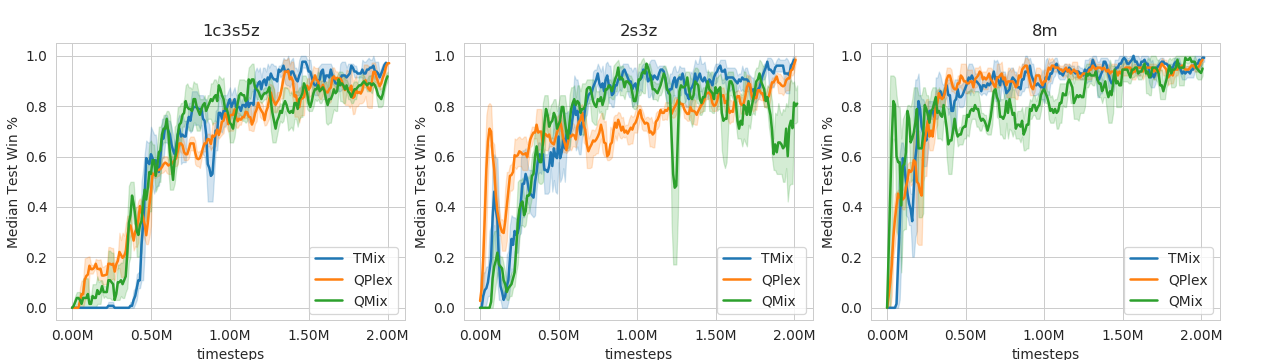}
    \includegraphics[width=1\textwidth]{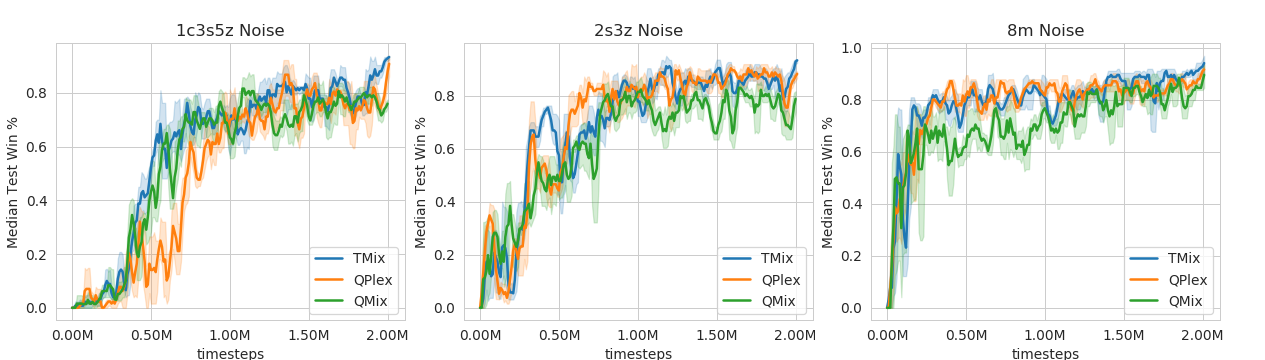}
    \caption{Win rate comparison between TransMix (labeled as TMix), QPlex, and QMIX. The top row represents win rate with regular global states while bottom row represents win rate with noisy global states.}
    \label{fig: Win_rate}
\end{figure*}

This paper also examines whether the usage of the transformer makes the MARL agents more robust to fog of war by injecting noise into the global state~\cite{fogofwar}. For these experiments, we add Gaussian noise, $\mathcal{N}(0, 0.05)$, to global states and train our model, QMix and QPlex from scratch. We follow the same training approach as discussed earlier and record the win rate. We observe that our method is less prone to noisy states compared to QMIX and QPlex, and that the performance drop of our method is less significant compared to others on the same maps without noisy states. These results are reported in Fig. \ref{fig: Win_rate} and Table \ref{table:3}. As the complexity of the map grows, the performance drop becomes more significant. 

\begingroup
\renewcommand{\arraystretch}{1.25}
\begin{table}[h]
\centering

\begin{tabularx}{0.4\textwidth} { 
  >{\raggedright\arraybackslash}X 
  | >{\centering\arraybackslash}X 
  | >{\centering\arraybackslash}X
  | >{\centering\arraybackslash}X}
 
 \centering{\textbf{Maps}} & {\textbf{TransMix}} & {\textbf{QMIX}} & {\textbf{QPlex}} \\ 
 \hline
 \hline
 3m & \textbf{99.82} & 95.31 & 97.41 \\ 
 8m & \textbf{96.87} & 93.75  & 93.75 \\
 2s3z & \textbf{93.75} & 81.67 & 89.15 \\
 1c3s5z & \textbf{93.75} & 78 & 83.88 \\
 3s5z & \textbf{84.18} & 65.76 & 72.62 \\
 \hline
\end{tabularx}
\caption{Noisy global state evaluation.  TransMix conclusively outperforms QMIX and QPlex when the global states are corrupted by noise.}
\label{table:2}
\end{table}
\endgroup

\begingroup
\renewcommand{\arraystretch}{1}
\begin{table}[!h]
\centering

\begin{tabularx}{0.45\textwidth} { 
  >{\raggedright\arraybackslash}X 
  | >{\centering\arraybackslash}X 
  | >{\centering\arraybackslash}X}
 
 \centering\textbf{Map} & \textbf{Ally Unit} & \textbf{Enemy Unit} \\
 \hline
 \hline
 2m\_vs\_1z & 2 Marines & 1 Zealot \\
 3m & 3 Marines & 3 Marines \\
 8m & 8 Marines & 8 Marines \\
 2s3z & 2 Stalkers and 3 Zealots & 2 Stalkers and 3 Zealots \\
 1c3s5z & 1 Colossi, 3 Stalkers and 5 Zealots & 1 Colossi, 3 Stalkers and 5 Zealots \\
 MMM & 1 Medivac, 2 Marauders and 7 Marines & 1 Medivac, 2 Marauders and 7 Marines \\
 bane\_vs\_bane & 20 Zerglings and 4 Banelings & 20 Zerglings and 4 Banelings \\
 \hline
 3s\_vs\_5z & 3 Stalkers & 5 Zealots \\
 3s5z & 3 Stalkers and 5 Zealots & 3 Stalkers and 5 Zealots \\
 5m\_vs\_6m & 5 Marines & 6 Marines \\
 8m\_vs\_9m & 8 Marines & 9 Marines \\
 2c\_vs\_64zg & 2 Colossi & 64 Zerglings \\
 10m\_vs\_11m & 10 Marines & 11 Marines \\
 \hline
\end{tabularx}
\caption{Details of SMAC Scenarios}
\label{table:3}
\end{table}
\endgroup

\section{Discussion}
The key to our method, TransMix, is the ability to learn the complex global and local contextual interaction amongst individual agents' Q-values, $Q_i$, the agents' histories, $h_i^t$, and the global states, $S_t$. The transformer correctly learns the context surrounding the correct choice of action, while remaining robust to noisy global states. 

Another aspect of our method is that the transformers are by design permutation invariant and thus do not depend on ordering. For instance, QMIX has to maintain the individual value function order which we do not require at all. While TransMix achieved a better test win rate on hard scenarios and comparable performance on easy scenarios, like other SMAC benchmarks, it performs poorly on super hard scenarios. 

The policies learnt by our method are very interesting. For instance, for the MMM map, the approach is aggressive. Each side has a heterogeneous team of 1 medivac, 2 mauraders, and 7 marines. The medivac agent remains behind all other agents while the marauders take the lead since they have heavy armors, and marines cover up the marauders.  This demonstrates that the transformer is good at learning the appropriate role mapping for different types of units.

\section{Conclusion}
This paper introduces TransMix, a value decomposition and mixing network for cooperative MARL tasks. TransMix uses a stack of transformer encoder layers trained in an end-to-end way, learning in a centralized fashion while executing the learned policies in a totally decentralized fashion. Our method is capable of representing non-linear value decomposition functions while maintaining consistency.

Results show that our method always outperforms QMIX and bests QPLEX on the majority of the hard StarCraft II Multi-agent Challenge scenarios. Furthermore, TransMix can still achieve good performance when the global states are perturbed with Gaussian noise, unlike QMIX.   In future work we seek to improve the performance on super hard scenarios by improving the exploration policy.

\section{Acknowledgments}
Research was partially sponsored by the Army Research Office and was accomplished under Cooperative Agreement Number W911NF-21-2-0103. The views and conclusions contained in this document are those of the authors and should not be interpreted as representing the official policies, either expressed or implied, of the Army Research Office or the U.S. Government.  The U.S. Government is authorized to reproduce and distribute reprints for Government purposes notwithstanding any copyright notation herein.

\bibliography{references}
\end{document}